\documentclass[twocolumn,prl,aps,floatfix,nopacs,superscriptaddress]{revtex4}
\usepackage{float}
\usepackage{xcolor}
\usepackage{graphicx}
\usepackage{hyperref}
\usepackage{amsmath}
\hypersetup{backref,pdfpagemode=FullScreen,colorlinks=true}
\hypersetup{
  colorlinks=ture,
  linkcolor=blue,
  urlcolor=blue,
  citecolor=blue,
  }

\begin{document}

\title{Inducing metal-insulator transition via disorder in correlated kagome systems}
\author{Qingzhuo Duan}
\affiliation{School of Physics and Astronomy, and Key Laboratory of Multiscale Spin Physics (Ministry of Education), Beijing Normal University, Beijing 100875, China}

\author{Hongdao Zhuge}
\affiliation{School of Physics and Astronomy, and Key Laboratory of Multiscale Spin Physics (Ministry of Education), Beijing Normal University, Beijing 100875, China}

\author{Zixuan Jia}
\affiliation{School of Physics and Astronomy, and Key Laboratory of Multiscale Spin Physics (Ministry of Education), Beijing Normal University, Beijing 100875, China}

\author{Tianxing Ma}
\email{txma@bnu.edu.cn}
\affiliation{School of Physics and Astronomy, and Key Laboratory of Multiscale Spin Physics (Ministry of Education), Beijing Normal University, Beijing 100875, China}
\date{\today}

\begin{abstract}
The metal-insulator transition is often accompanied by fascinating quantum phenomena, including superconducting domes, antiferromagnetic phase transitions, and quantum spin liquids. Concurrently, kagome materials are predominantly metallic, necessitating the realization of insulating states to fully exploit their significant potential in logic and optoelectronic device applications. To address this, we investigate the electronic transport and magnetic properties in correlated kagome systems with hopping disorder using the determinant quantum Monte Carlo method. Through comprehensive analysis of the kinetic energy, dc conductivity, and density of states at the Fermi level, we demonstrate that the cooperative interplay between hopping disorder and electron correlations promotes electron localization. Within the insulator, an increase in the disorder level reduces the Coulomb interaction required for the Mott transition. Additionally, while disorder partially suppresses antiferromagnetic ordering, it remains insufficient to induce a complete magnetic transition. Finally, we summarize two schematic regions distinguishing between antiferromagnetic metal, correlated Anderson insulator, and disordered Mott insulator. Our study advances the understanding of metal-insulator transition in kagome systems by disorder and provides actionable insights for experimental control of these transitions.
\end{abstract}

\maketitle

\section{I. INTRODUCTION}
\vspace{-0.2cm}

The kagome lattice, a geometrically frustrated corner-sharing triangular network, initially attracted extensive attention in the study of quantum spin liquids, and has subsequently emerged as an ideal platform for investigating electron correlation effects, including metal-insulator transition (MIT), magnetism, unconventional superconductivity, and nontrivial topological states \cite{PhysRevB.102.224415,PhysRevLett.127.236401,PhysRevLett.134.086902,nature08917,Nature.11659,science.aab2120,PhysRevLett.126.247001,ZHU2023157817,FANG2024112725,Yang_2024}. In real materials, most kagome compounds are gapless metals \cite{s41467-025-56582-7,PhysRevLett.125.247002,s41567-023-01985-w}, a severe limitation that restricts their applications in logic and optoelectronic devices \cite{FANG2024112725,ZHU2023157817}. Therefore, studying the MIT in kagome systems holds significant theoretical importance and promising application prospects.

Three types of insulators can emerge following a MIT. Systems with completely filled valence bands are termed band insulators, where the fully occupied bands remain spatially symmetric under an applied electric field, resulting in zero current density \cite{PhysRevB.92.155312,PhysRevLett.83.2014}. Real materials typically possess structural defects, lattice distortions, and inhomogeneous external fields, which introduce disorder into the system; such disorder weakens coherent interference effects and influences quantum transport, thereby inducing Anderson localization and the corresponding insulating phase \cite{PhysRev.109.1492,RevModPhys.50.191,PhysRevLett.96.063904}. When electron correlations are considered, metallic systems can undergo a transition to an insulating state driven by the competition between the energy gap and kinetic energy; electrons in narrow bands near the Fermi level become localized, and the system becomes a Mott insulator \cite{PhysRevB.108.235163,PhysRevB.101.245161}. Recently, it has been found that breathing effects and anisotropy effectively modulate the critical correlation strength for the MIT in correlated kagome systems \cite{cpl_42_9_090712,PhysRevB.108.235163}. The underlying physical mechanisms of them can be attributed to the gap opening at the Fermi level due to lattice symmetry breaking, analogous to those in band insulators.

Over the past decades, the nature of the disorder-driven MIT in two-dimensional interacting systems has been intensively debated. Finkelstein \textit{et al.} \cite{BF01304171} first predicted the existence of metallic states at zero magnetic field, and subsequent studies \cite{PhysRevB.57.R9381} further confirmed the possibility of metallic behavior and the MIT. Through perturbative renormalization group analyses of the combined effects of interactions and disorder, a quantum critical point was identified separating two distinct phases: a metallic phase stabilized by electronic correlations and an insulating phase where disorder prevails over interactions \cite{science.1115660}. It is now widely accepted that understanding the MIT requires treating electron correlations and disorder on an equal footing, as both are invariably present in real materials \cite{Curro_2009,science.1107559}. When disorder and interactions are both strong, perturbative approaches typically break down \cite{Rev.66}. Although quantum Monte Carlo simulations may suffer from the sign problem, disorder may alleviate this issue \cite{PhysRevB.105.045132}.

Within the framework of quantum Monte Carlo methods, intriguing the MIT have been reported in various physical systems \cite{PhysRevLett.120.116601,PhysRevB.105.045132}. By investigating the disordered Hubbard model on a square lattice at $\frac{1}{4}$ filling, it was demonstrated that repulsive interactions between electrons can significantly enhance conductivity in two-dimensional systems incorporating both interactions and disorder, providing evidence for the transition \cite{PhysRevLett.83.4610}. In two-dimensional honeycomb lattice systems featuring a density of states that vanishes linearly at the Fermi level, a novel disorder-induced nonmagnetic insulator was found to emerge from the zero-temperature quantum critical point, separating a semimetal from a Mott insulator \cite{Matter.12}. Furthermore, upon doping the disordered honeycomb lattice, the critical disorder strength required for the MIT decreases \cite{PhysRevB.109.045107}.

Motivated by the aforementioned studies, we investigate the Hubbard model on a disordered kagome lattice using the determinant quantum Monte Carlo (DQMC) method. We find that the introduction of hopping disorder at stronger correlation strengths can alleviate the sign problem, whereas it has the opposite effect at weaker correlations. In the disorder-free system, we obtain a critical interaction strength $U_c \approx 6.4$ for the MIT, in excellent agreement with previous high-precision studies \cite{PhysRevB.93.245123,PhysRevB.107.035134}. At weak coupling ($U \lesssim 4$), a disorder strength of $\Delta \approx 3$ is required to drive the system from an antiferromagnetic (AFM) metal into a gapless Anderson insulator. As the Coulomb repulsion increases, $\Delta_\mathrm{c}$ required for the MIT decreases, indicating that the interplay between disorder and interactions facilitates more efficient electron localization. Within the disordered insulator, there is a transition from the Anderson-like insulator to the Mott-like insulator. Furthermore, although disorder partially suppresses AFM order, it is insufficient to induce a magnetic transition.

\vspace{-0.2cm}  
\section{II. MODEL AND METHOD}
\vspace{-0.3cm}

The Hamiltonian for disordered Hubbard model on a kagome lattice is defined as

\begin{equation}
\hat{H} = -\sum_{\langle i,j \rangle, \sigma} t_{ij} (\hat{c}_{i\sigma}^{\dagger} \hat{c}_{j\sigma} + \hat{c}_{j\sigma}^{\dagger} \hat{c}_{i\sigma}) - \mu \sum_{i\sigma} \hat{n}_{i\sigma} + U \sum_{i} \hat{n}_{i\uparrow} \hat{n}_{i\downarrow}.
\label{Eq1}
\end{equation}

\begin{figure}[t]
	\centering
	\includegraphics[width=0.8\linewidth]{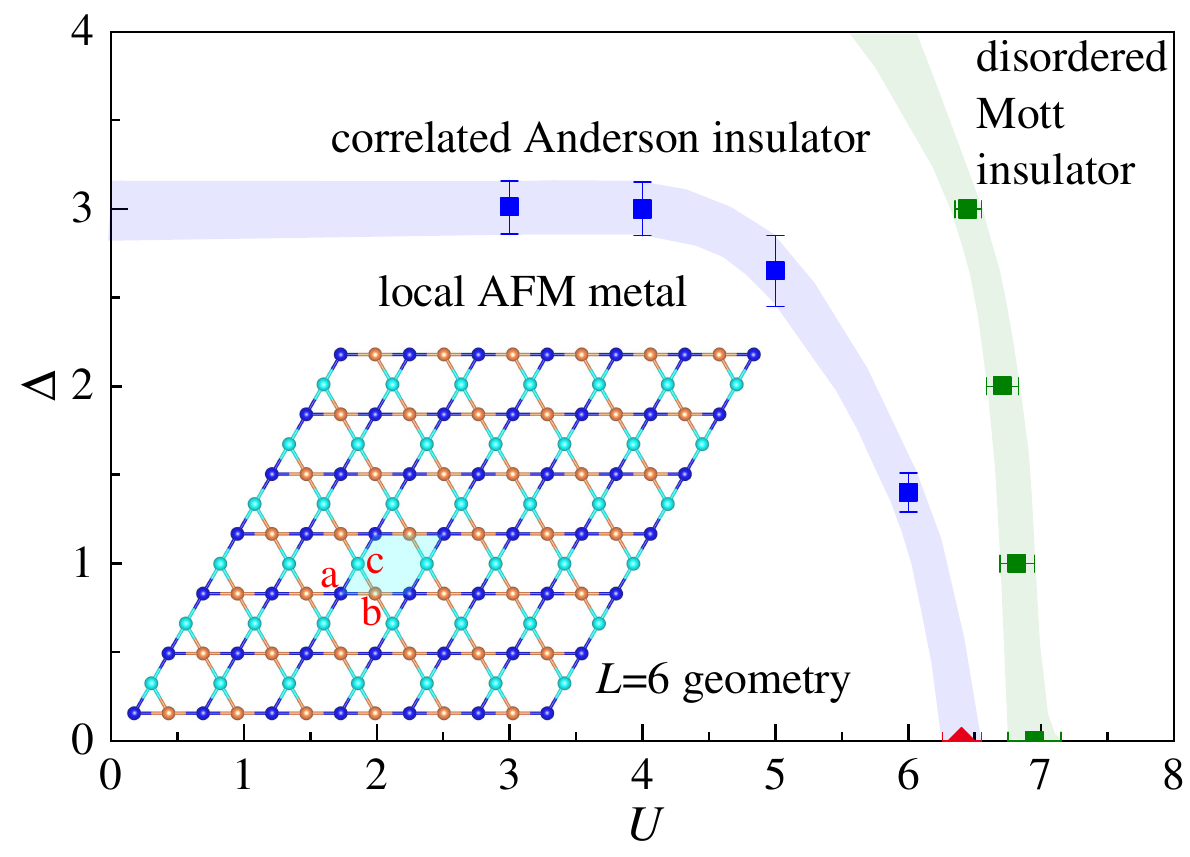}
	\caption{Phase diagram of the disordered Hubbard model on the kagome lattice at half filling. The blue shaded region correspond to the critical disorder strength $\Delta_{\mathrm{c}}$, which schematically separates the local AFM metal from the correlated Anderson insulator. The critical correlation strength $U_{\mathrm{c}}$ of the disorder-free system is marked by red diamonds. The $U_{\mathrm{c}}$ for distinguishing between correlated Anderson insulator and disordered Mott insulator is represented by the green shaded region. The inset shows the geometry of the $L=6$ lattice. The sublattices are labeled by a, b, and c cites. The primitive cell is highlighted by the transparent cyan region.}
	\label{Fig1}
\end{figure}

\noindent Here, $t_{ij}$ represent the hopping amplitude between two nearest-neighbor sites $i$ and $j$, $\hat{c}_{i\sigma}^{\dagger}$ ($\hat{c}_{j\sigma}$) is the creation (annihilation) operator at site $i$ ($j$) with spin $\sigma$, and $\hat{n}_{i\sigma} = \hat{c}_{i\sigma}^{\dagger}\hat{c}_{i\sigma}$ is the number operator with spin $\sigma$ at site $i$. $\mu$ is the chemical potential and determines the electron density of the system. $U$ represents the on-site repulsive interaction.
Hopping disorder is induced by modifying the matrix element $t_{ij}$ of the hopping matrix, which is chosen from $t_{ij} \in [t - \Delta/2, t + \Delta/2]$ and zero otherwise with a probability $P(t_{ij}) = 1/\Delta$. We set $t = 1$ as the reference energy scale. The disorder strength is characterized by $\Delta$, which quantifies the fluctuation amplitude of the hopping matrix elements $t_{ij}$. To obtain statistically reliable results in the presence of disorder, we perform an average over 20 disorder realizations, a procedure that has been demonstrated to effectively suppress statistical errors arising from randomness \cite{PhysRevLett.83.4610,PhysRevLett.120.116601,PhysRevB.105.045132,PhysRevB.109.045107}.
The geometry of the simulated system is illustrated in \hyperref[Fig1]{Fig. 1}. All calculations are performed on a $3 \times L^2$ kagome cluster with linear size $L = 6$.

We employ the DQMC method \cite{PhysRevD.24.2278,PhysRevB.40.506} to numerically investigate the MIT in the Hubbard model defined by \hyperref[Eq1]{Eq. 1}. As a nonperturbative approach, DQMC provides an exact numerical technique for studying the Hubbard model at finite temperatures. The method proceeds by expressing the partition function $Z = \mathrm{Tr}\, e^{-\beta \hat{H}}$ as a path integral, which is then discretized into $\Delta\tau$ time slices along the imaginary time axis $(0, \beta)$. Here the inverse temperature is $\beta \equiv 1/(k_{\mathrm{B}}T)$, with $k_{\mathrm{B}}$ denoting the Boltzmann constant. While the kinetic energy term is already quadratic, the on-site interaction is decoupled into a quadratic form via a discrete Hubbard-Stratonovich (HS) transformation. Subsequently, the partition function can be analytically transformed into the product of two fermion determinants by integrating the quadratic Hamiltonian term. These determinants correspond to the spin-up and spin-down states, respectively. The Metropolis algorithm is employed for efficient sampling, with the time-step $\Delta\tau = 0.1$ chosen to ensure sufficiently small Trotter errors. To assess the reliability of our results, we monitor the average fermion sign $\langle \mathrm{sign} \rangle$ as a diagnostic of the severity of the sign problem:

\begin{equation}
\langle \mathrm{sign} \rangle = \frac{\sum_{\mathcal{X}} \det M_{\uparrow}(\mathcal{X}) \det M_{\downarrow}(\mathcal{X})}{\sum_{\mathcal{X}} |\det M_{\uparrow}(\mathcal{X}) \det M_{\downarrow}(\mathcal{X})|},
\label{Eq2}
\end{equation}

\noindent where $\mathcal{X}$ represents the HS configurations, comprising spatial sites and imaginary time slices. The matrix $M_{\sigma}(\mathcal{X})$ corresponds to spin species $\sigma$. A value of $\langle \mathrm{sign} \rangle = 1$ indicates the absence of the sign problem.

To characterize the MIT in the disordered kagome lattice, we compute the temperature-dependent dc conductivity $\sigma_{\mathrm{dc}}(T)$. This quantity is extracted from the current-current correlation function $\Lambda_{xx}(\boldsymbol{q},\tau)$ in momentum $\boldsymbol{q}$ and imaginary time $\tau$, defined as~\cite{PhysRevLett.83.4610,PhysRevB.54.R3756}

\begin{equation}
\sigma_{\mathrm{dc}}(T) = \frac{\beta^2}{\pi} \Lambda_{xx}\!\left(\boldsymbol{q}=0, \tau=\frac{\beta}{2}\right).
\label{Eq3}
\end{equation}

\noindent Here, $\beta = 1/T$ (we set $k_{\mathrm{B}} = 1$), and 

\begin{equation}
\Lambda_{xx}(\boldsymbol{q},\tau) = \langle j_x(\boldsymbol{q},\tau) j_x(-\boldsymbol{q},0) \rangle,
\label{Eq4}
\end{equation}

\noindent where $j_x(\boldsymbol{q},\tau)$ denotes the Fourier transform of the unequal-time current-current correlation function,

\begin{equation}
j_x(\boldsymbol{q},\tau) = e^{H\tau} \left[ \mathrm{i} \sum_{\sigma} t_{\boldsymbol{q}+\boldsymbol{x},\boldsymbol{q}} (c_{\boldsymbol{q}+\boldsymbol{x},\sigma}^{\dagger} c_{\boldsymbol{q},\sigma} - c_{\boldsymbol{q},\sigma}^{\dagger} c_{\boldsymbol{q}+\boldsymbol{x},\sigma}) \right] e^{-H\tau}.
\label{Eq5}
\end{equation}

\noindent For disordered systems, the \hyperref[Eq3]{Eq. 3} provides a good approximation if the temperature is lower than the energy scale set by the disorder strength $\Delta$. Additionally, The \hyperref[Eq3]{Eq. 3} has been extensively validated in previous studies of the Hubbard model~\cite{PhysRevLett.83.4610,PhysRevB.54.R3756,PhysRevB.107.035134,PhysRevB.109.045107,PhysRevLett.120.116601}.

\begin{figure}[t]
	\centering
	\includegraphics[width=\linewidth]{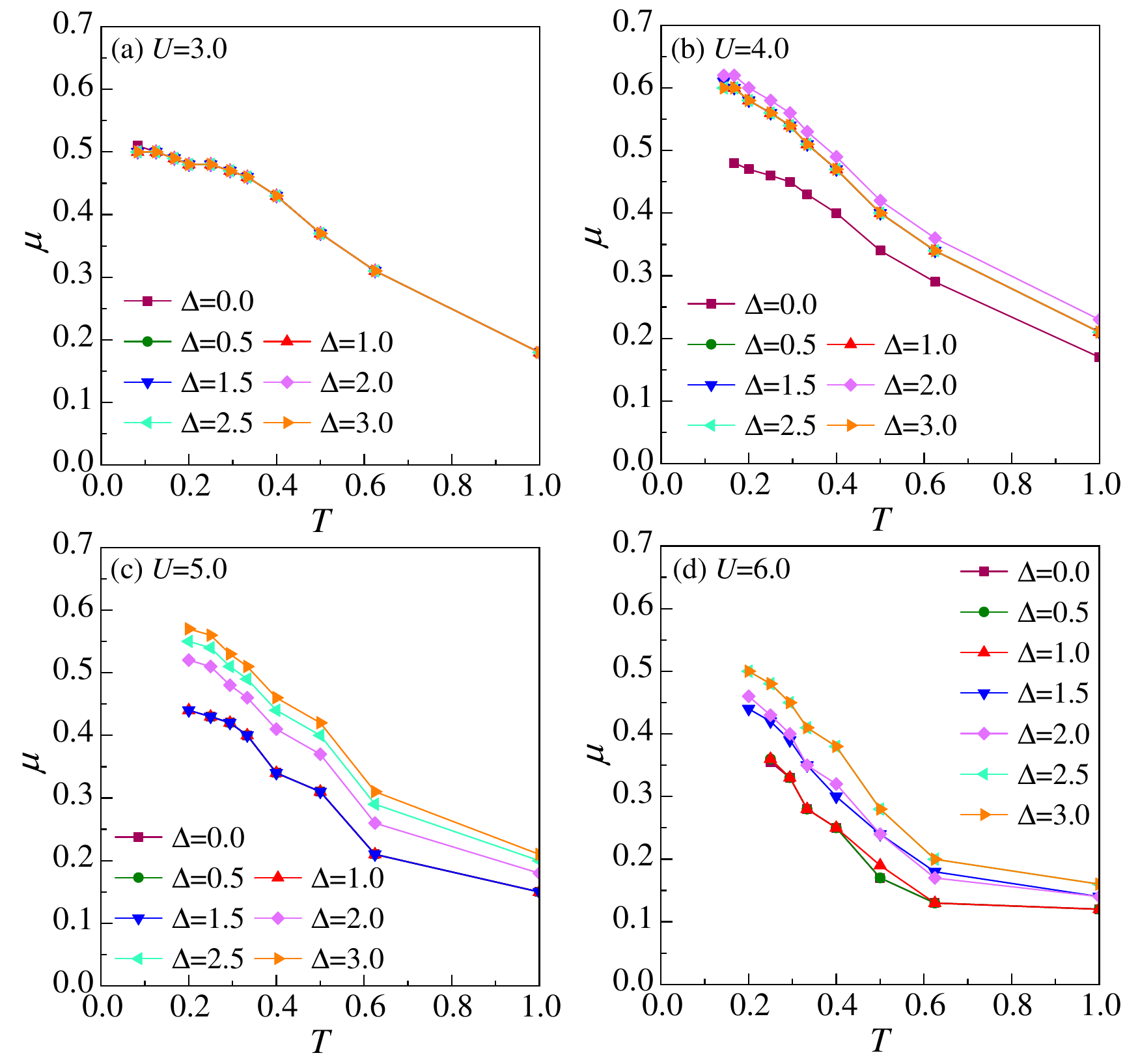}
	\caption{Chemical potential $\mu$ that leads to the half filling of the system as a function of temperature $T$ for various correlation strength $U$ and disorder strength $\Delta$. The data regarding $\Delta=0.0$ are adopted from our previous work~\cite{cpl_42_9_090712}.}
	\label{Fig2}
\end{figure}

We also evaluate the kinetic energy per site, obtained by averaging the hopping terms:
\begin{equation}
\langle K \rangle = -\frac{1}{N} \sum_{\langle i,j \rangle, \sigma} t_{ij} (\hat{c}_{i\sigma}^{\dagger} \hat{c}_{j\sigma} + \hat{c}_{j\sigma}^{\dagger} \hat{c}_{i\sigma}).
\end{equation}

\noindent To circumvent the need for numerically analytical continuation, we evaluate the imaginary-time Green's function at $\tau = \beta/2$. This quantity provides an estimate of the low-energy spectral weight near the Fermi level and is widely used in DQMC simulations as a proxy to distinguish metallic from insulating behavior. Specifically, we focus on the density of states at the Fermi level,
\begin{equation}
N(0) \approx \beta \times G(\boldsymbol{r}=0, \tau=\beta/2),
\end{equation}
\noindent where $G(\boldsymbol{r},\tau)$ denotes the imaginary-time-dependent Green's function.

To probe the magnetic properties of the disordered kagome lattice, we compute the real-space spin-spin correlation functions~\cite{PhysRevB.59.3321,PhysRevB.108.235163}

\begin{equation}
c^{\alpha\beta}(\boldsymbol{r}) = \frac{1}{3} \langle \boldsymbol{S}_{\boldsymbol{r}_0}^{\alpha} \cdot \boldsymbol{S}_{\boldsymbol{r}_0+\boldsymbol{r}}^{\beta} \rangle,
\end{equation}

\noindent with $\boldsymbol{r}_0$ denoting a reference lattice site and $\boldsymbol{r}$ the relative displacement between sites. The indices $\alpha$ and $\beta$ label the sublattices $a$, $b$, or $c$, which are marked in \hyperref[Fig1]{Fig. 1}.

\vspace{-0.2cm}  
\section{III. RESULT AND DISCUSSION}
\vspace{-0.3cm}

While DQMC provides an unbiased framework for studying correlated electron systems, it suffers from the notorious fermion sign problem, which leads to exponentially increasing statistical noise at low temperatures \cite{PhysRevB.41.9301,PhysRevLett.94.170201}. This complication is absent in half-filled systems with particle-hole symmetry (PHS), such as bipartite lattices including the square and honeycomb structures. However, the kagome lattice is non-bipartite and lacks PHS at any filling, giving rise to a severe sign problem whose severity depends on system size, temperature, and interaction strength. In this work, we focus on the half-filled regime with average electron density $n = 1$ per site, a parameter region of considerable physical interest. \hyperref[Fig2]{Figure 2} illustrates how $\mu$ required to maintain half-filling varies with temperature for different values of $U$ and $\Delta$. It can be observed that as the temperature increases, the chemical potential \(\mu\) required to maintain half-filling decreases continuously. This behavior arises because thermal excitation drives electrons near the Fermi surface to higher energy regions, where the density of states is typically larger than that at lower energies, leading to an increase in occupation. Consequently, reducing \(\mu\) helps preserve the total filling at half capacity.

\begin{figure}[t]
	\centering
	\includegraphics[width=\linewidth]{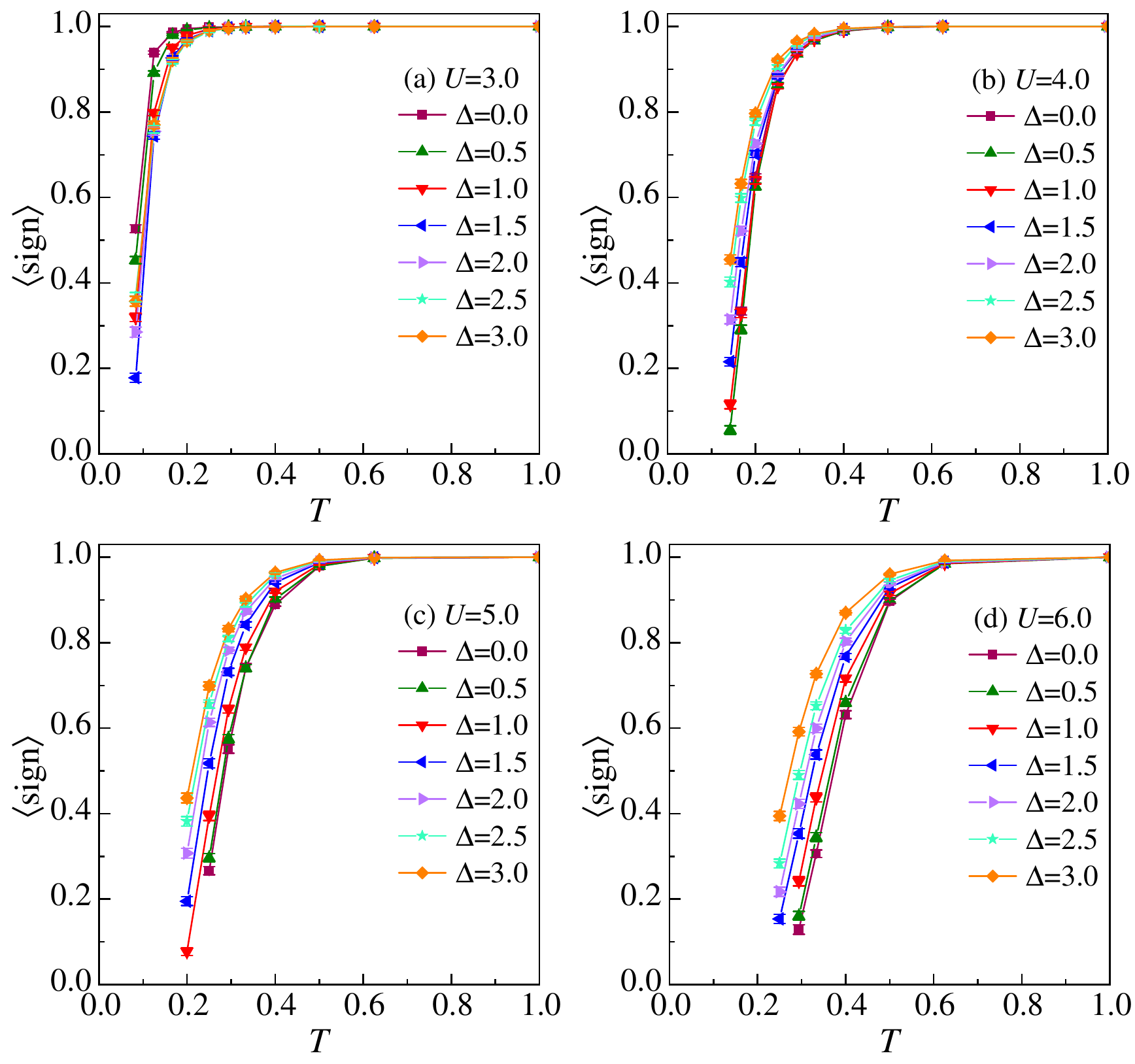}
	\caption{Average sign $\langle \mathrm{sign} \rangle$ as a function of temperature $T$ for various correlation strength $U$ and disorder strength $\Delta$. The data regarding $\Delta=0.0$ are adopted from our previous work~\cite{cpl_42_9_090712}.}
	\label{Fig3}
\end{figure}

To ensure the reliability of the results, we computed $\langle \mathrm{sign} \rangle$ as a function of temperature under different values of $U$ and $\Delta$, as shown in \hyperref[Fig3]{Figure 3}. When $U$ = 4, 5, and 6, introducing disorder can effectively alleviate the sign problem, and the degree of alleviation increases with larger $U$ (see \hyperref[Fig3]{Fig. 3(b)-(d)}). However, for $U$ = 3, introducing disorder instead exacerbates the sign problem, contrary to the expectation that disorder suppresses magnetic order and thus mitigates the sign problem (see \hyperref[Fig3]{Fig. 3(a)}). One possible explanation is that at strong interaction strengths, disorder alleviates the sign problem by suppressing long-range antiferromagnetic order and reducing magnetic correlations (see \hyperref[Fig8]{Fig. 8(d)}). However, at weak interaction strengths, the sign problem is less affected by magnetic ordering. In this regime, disorder leads to a random spatial distribution of single-particle states, causing the sign of the weight for each configuration in the Monte Carlo sampling to fluctuate more strongly, thereby compromising the positive definiteness of the weights.

\begin{figure}[t]
	\centering
	\includegraphics[width=\linewidth]{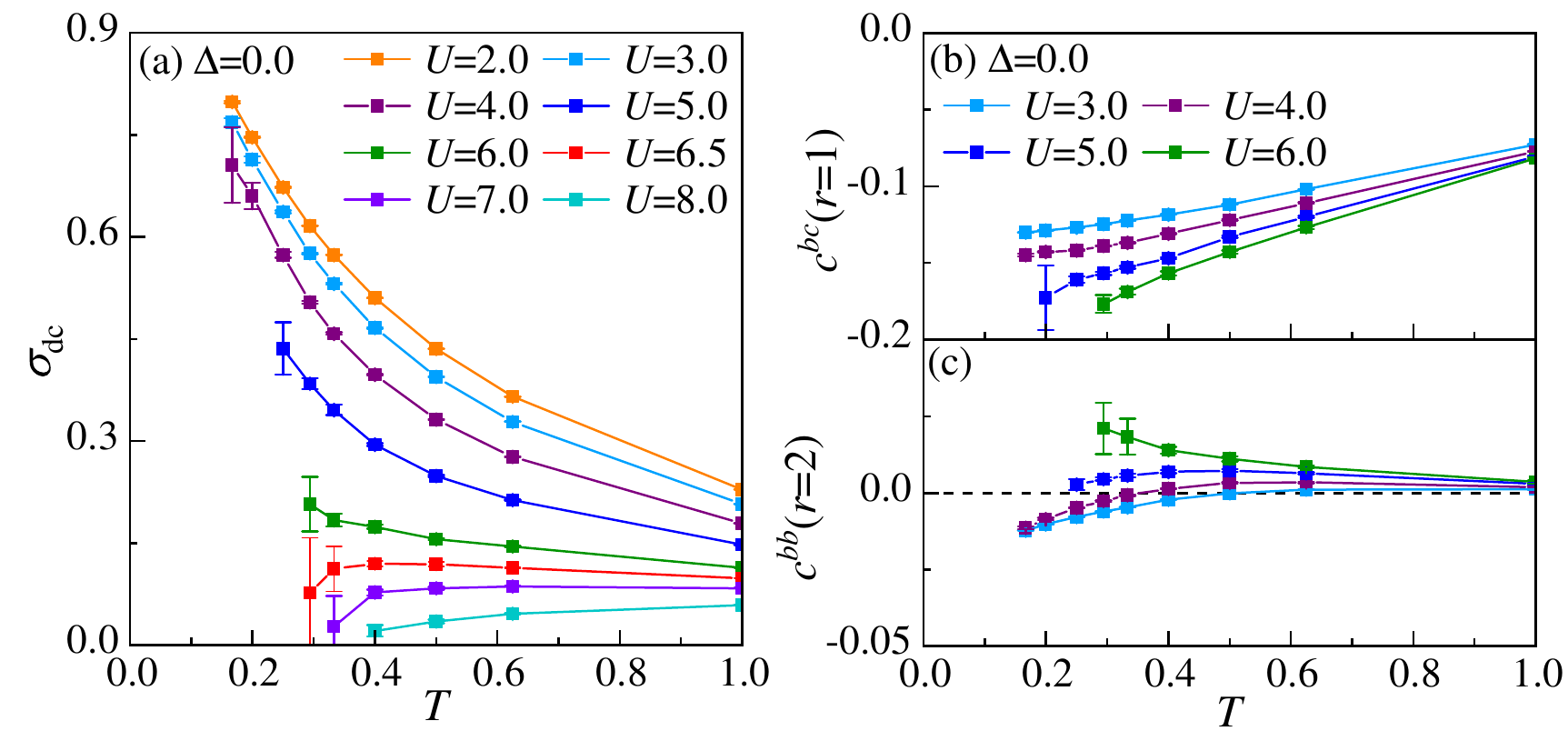}
	\caption{(a) dc conductivity $\sigma_{\mathrm{dc}}$, (b) nearest-neighbor spin-spin correlation function between $b$ and $c$ sites $c^{bc}(r=1)$ and (c) next-nearest-neighbor correlation between $b$ sites $c^{bb}(r=2)$ as a function of temperature $T$ for various correlation strength $U$ in kagome system. Part of the data are adopted from our previous work \cite{cpl_42_9_090712}. The legend in (c) is the same as that in (b).}
	\label{Fig4}
\end{figure}

\begin{figure}[t]
	\centering
	\includegraphics[width=\linewidth]{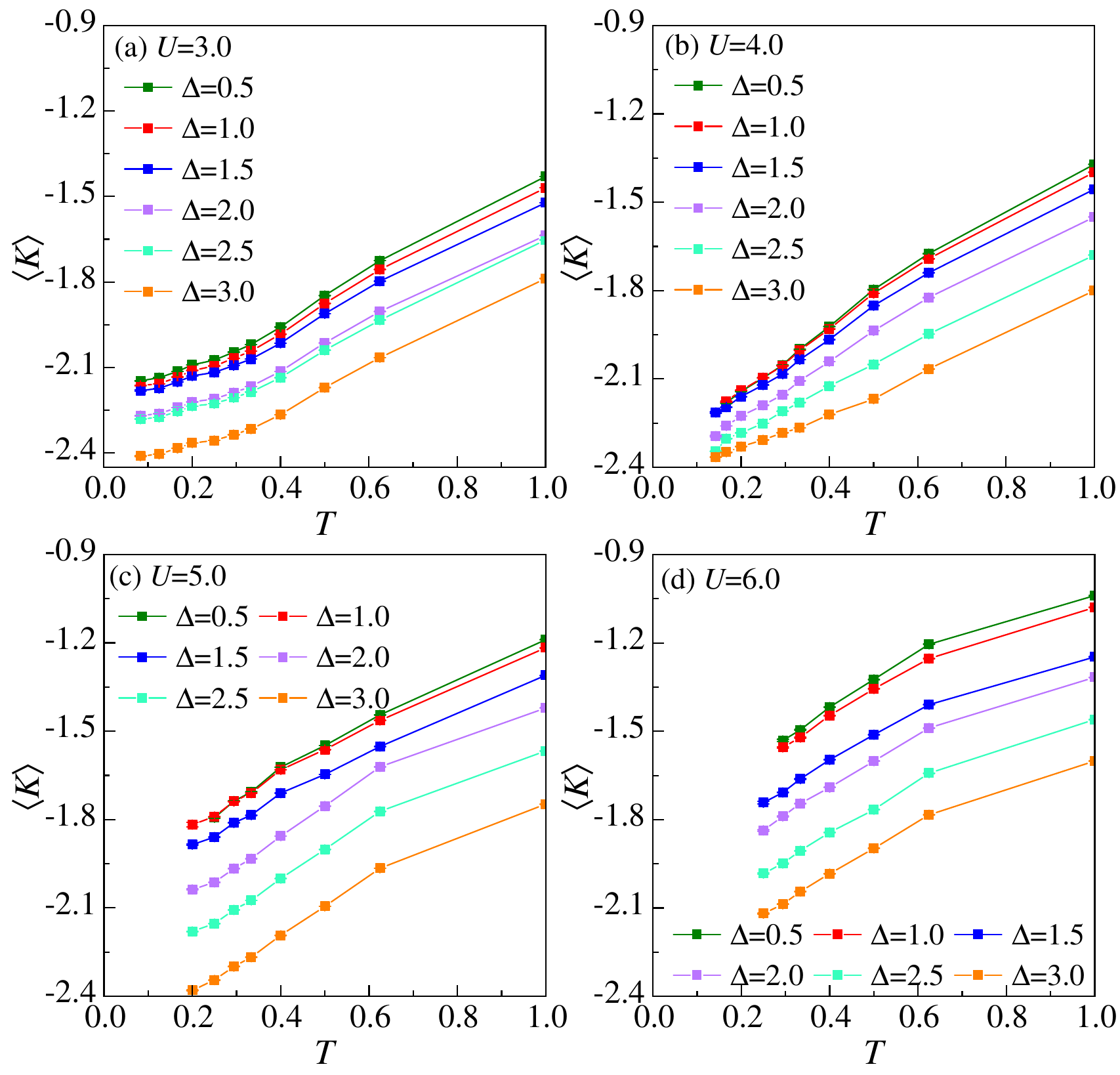}
	\caption{Kinetic energy per site $\langle K \rangle$ as a function of temperature $T$ for various correlation strength $U$ and disorder strength $\Delta$.}
	\label{Fig5}
\end{figure}

Next, we present results for the disorder-free system. Part of the data are adopted from our previous work \cite{cpl_42_9_090712}. \hyperref[Fig4]{Fig. 4(a)} shows $\sigma_{\mathrm{dc}}(T)$ computed for several representative values of $U$. For $U < 6.5$, we find $\partial\sigma_{\mathrm{dc}}/\partial T < 0$ as $T$ increases, with $\sigma_{\mathrm{dc}}(T)$ diverging as $T \to 0$, indicative of metallic behavior. Conversely, for $U \geq 6.5$, $\partial\sigma_{\mathrm{dc}}/\partial T > 0$ at low temperatures, signaling fermion localization. This qualitative change demonstrates a MIT \cite{PhysRevLett.83.4610}. From this analysis, we extract the critical interaction strength for the MIT as $U_\mathrm{c} = 6.40 \pm 0.13$, which is consistent with the results of previous high-precision studies \cite{PhysRevB.93.245123,PhysRevB.107.035134}. This transition point is indicated in \hyperref[Fig1]{Figure 1}. To characterize the magnetic transition, \hyperref[Fig4]{Fig. 4(b)} displays the nearest-neighbor spin-spin correlation function $c^{bc}(r=1)$ between $b$ and $c$ sites as a function of temperature for various $U$. For all $U$, $c^{bc}(r=1)$ decreases monotonically from negative values upon cooling. At fixed $T$, the magnitude $|c^{bc}(r=1)|$ grows with increasing $U$, indicating that stronger electron repulsion promotes localization and thereby enhances magnetic ordering. The next-nearest-neighbor correlation $c^{bb}(r=2)$ between $b$ sites is shown in \hyperref[Fig4]{Fig. 4(c)}. $c^{bb}(r=2)$ is at least one order of magnitude smaller than $c^{bc}(r=1)$. At $U = 3$ and $U = 4$, $c^{bb}(r=2)$ exhibits small negative values. As $U$ increases to 5 and 6, $c^{bb}(r=2)$ gradually develops a weak ferromagnetic (FM) character. These observations indicate that the disorder-free kagome lattice hosts strong short-range AFM correlations, which are enhanced with increasing $U$. This conclusion is consistent with many experimentally discovered kagome materials, such as $\rm ZnCu_3(OH)_6Cl_2$, $\rm FeGe$, and $\rm YMn_6Sn_6$ \cite{Nature.022.05034,PhysRevLett.98.107204,sciadv.abe2680}.

\begin{figure}[t]
	\centering
	\includegraphics[width=\linewidth]{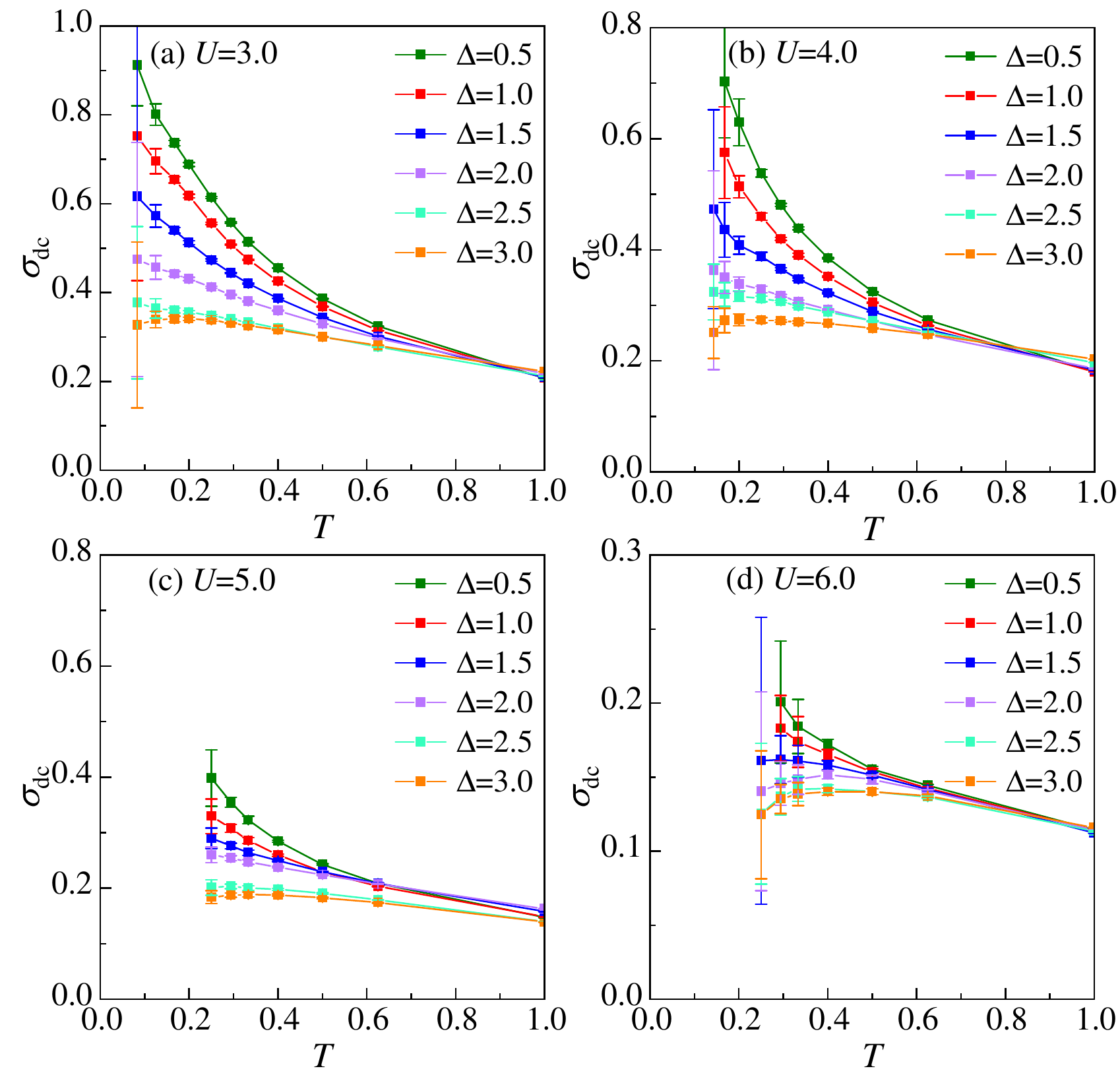}
	\caption{dc conductivity $\sigma_{\mathrm{dc}}$ as a function of temperature $T$ for various correlation strength $U$ and disorder strength $\Delta$.}
	\label{Fig6}
\end{figure}

Next we move on to the disordered case, first examining the kinetic energy. \hyperref[Fig5]{Figure 5} displays the kinetic energy per site $\langle K \rangle$ for the kagome lattice as a function of temperature for various $\Delta$ and $U$. For all $U$, the magnitude of $\langle K \rangle$ ($|\langle K \rangle|$) gradually increases with increasing $\Delta$.
We attempt to provide a qualitative interpretation of this phenomenon: disorder may induce a reconstruction of the Fermi surface, which could be accompanied by a more dispersed momentum distribution of electrons. Near the Fermi surface, the kinetic energy of electrons may become constrained, driving the system toward an insulating tendency. However, the kinetic energy of electrons in other regions may be enhanced, leading to a higher average kinetic energy overall.
\hyperref[Fig5]{Fig. 5(a)} indicates that for $U=3$ and $T < 0.4$, the rate of increase of $|\langle K \rangle|$ with decreasing temperature slows down. However, even down to $T=0.1833$, no transition point satisfying $\partial\langle K\rangle/\partial T < 0$ is observed, indicating that a MIT does not occur. For $U = 4, 5, 6$, the condition $\partial\langle K\rangle/\partial T < 0$ is also not satisfied (see \hyperref[Fig5]{Fig. 5(b)-(d)}). Consequently, $\langle K \rangle$ does not capture the MIT in this system.

Given the above, other quantities should be examined to locate the transition behavior. We therefore investigate the dc conductivity $\sigma_{\mathrm{dc}}$ as a function of temperature $T$, calculated for several representative values of $U$ and $\Delta$ (as shown in \hyperref[Fig6]{Figure 6}). \hyperref[Fig6]{Fig. 6(d)} shows that for \( U = 6 \) and \( \Delta < 1.5 \), the condition \(\partial\sigma_{\mathrm{dc}}/\partial T < 0\) is satisfied as \(T\) increases, with \(\sigma_{\mathrm{dc}}(T)\) diverging as \(T \to 0\), indicating metallic behavior. However, when \(\Delta \geq 1.5\), \(\partial\sigma_{\mathrm{dc}}/\partial T \geq 0\) at low temperatures, signaling a transition to an insulator. From this analysis, we extract critical disorder strength \(\Delta_{\mathrm{c}} = 1.40 \pm 0.11\) for \(U = 6\). \hyperref[Fig6]{Fig. 6(c)} displays \(\Delta_{\mathrm{c}}= 2.65 \pm 0.20\) when \( U = 5 \). \hyperref[Fig6]{Fig. 6(a)-(b)} show that $\Delta_{\mathrm{c}}=3.00 \pm 0.15$ for both $U=3$ and $U=4$, indicating that $\Delta_{\mathrm{c}}$ likely saturates when $U \leq 4$. Those indicate that the introduction of electron correlations will reduce $\Delta_{\mathrm{c}}$.
Based on these results, we employ schematic shaded region to differentiate the metallic and insulating states in \hyperref[Fig1]{Figure~1}. We acknowledge that, given the finite system sizes and the finite temperature window accessible in our DQMC simulations, the observed MIT should be interpreted as a finite-temperature crossover. In a disordered correlated system, spatial inhomogeneity naturally smears any would-be sharp transition into a broad crossover, which is physically expected and experimentally relevant.

\begin{figure}[t]
	\centering
	\includegraphics[width=\linewidth]{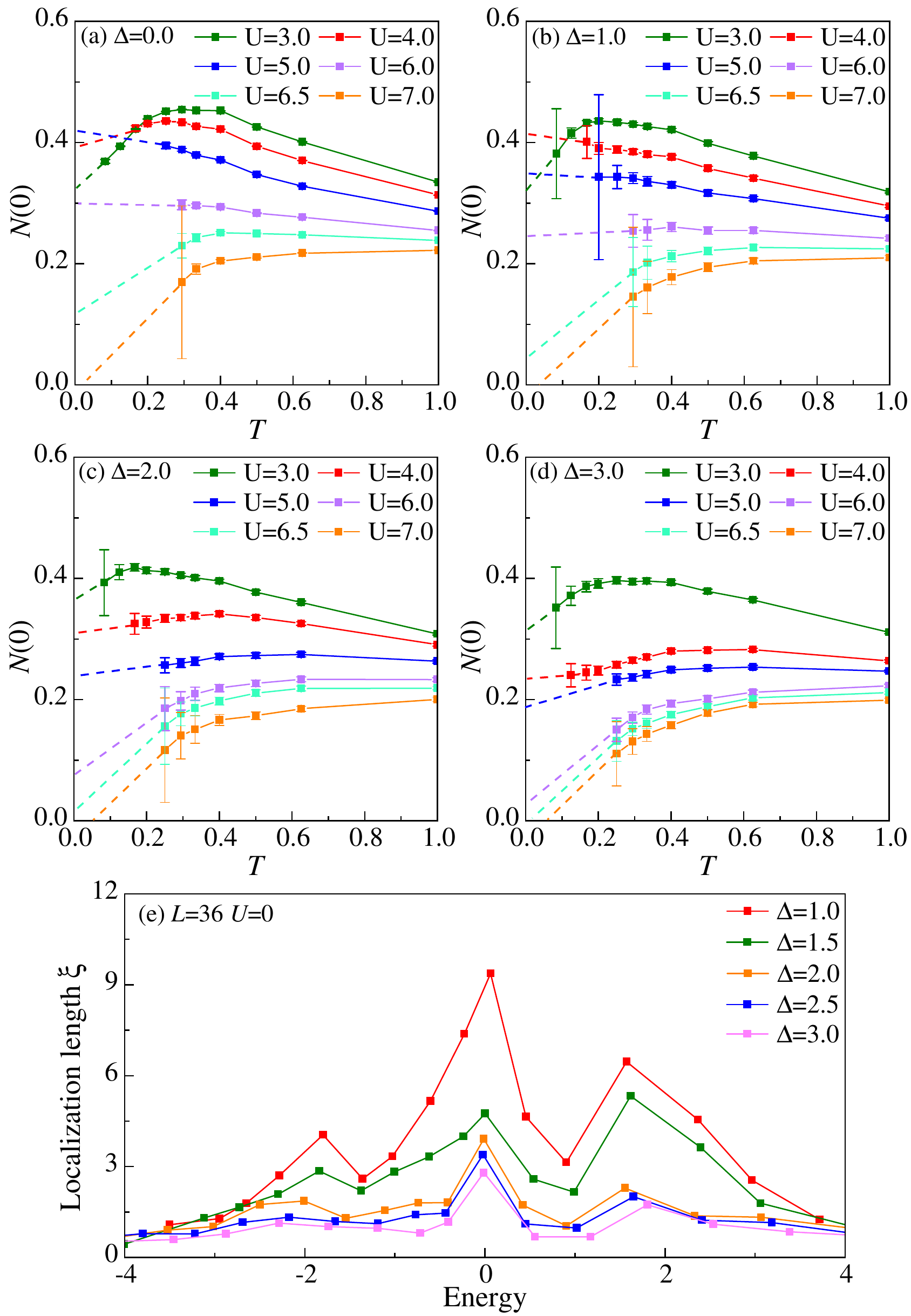}
	\caption{(a)-(d) Density of states at the Fermi level $N(0)$ as a function of temperature $T$ for various correlation strength $U$ and disorder strength $\Delta$. Since calculations at even lower temperatures are challenging, our fits in the $T \to 0$  limit (dashed lines) are to be interpreted on a qualitative level. (e) Localization length $\xi$ as a function of energy for different disorder strengths at $L=36$, $U=0$.}
	\label{Fig7}
\end{figure}

The presence or absence of a single-particle gap in the insulating state is a fundamental electronic property of considerable interest. In the absence of disorder, a sufficiently large Coulomb interaction opens a Mott gap in the excitation spectrum \cite{PhysRevB.72.085123}. Conversely, for non-interacting systems, disorder-induced Anderson insulators are gapless at the Fermi level in the thermodynamic limit \cite{RevModPhys.50.191}. Although the gap is not an order parameter associated with symmetry breaking, it nevertheless serves to establish the existence of a Mott insulator: in the thermodynamic limit, whether the density of states at the Fermi level $N(0)$ is zero serves as a critical criterion for distinguishing between gapped and gapless systems.
For the disorder-free system, \hyperref[Fig7]{Fig. 7(a)} reveals that our qualitative fit of $N(0)$ in the $T \to 0$ limit (dashed line in the figure) falls near zero and a gap may open. Here, we set the critical Coulomb interaction strength $U_\mathrm{c} \approx 6.95$.
When increase the level of randomness, \hyperref[Fig7]{Fig. 7(b)-(d)} indicate that the gap develops at progressively lower interaction strengths, with $U_\mathrm{c} \approx 6.82$, 6.71, and 6.45 at $\Delta = 1.0$, 2.0, and 3.0, respectively.

To enhance the reliability of our results, we examine the finite-size effects of $N(0)$ for $L=$5,6,7, and 8 (see \hyperref[Fig10]{Fig. 10(c)}). The values of $N(0)$ obtained for different \textit{L}, differ by less than 5$\%$, demonstrating that $N(0)$ has negligible finite-size dependence.
Moreover, to reveal the signatures of Anderson localization in the system, we calculate the Anderson localization length $\xi$ for a sufficiently large lattice with $L = 36$ and $U = 0$. In disordered physical systems, the eigenstates exhibit localized behavior and decay exponentially in space as $\psi(x) \sim e^{-x/\xi}$, where $\xi$ is the Anderson localization length measured in units of the lattice constant \cite{zbMATH07265614}. For $U = 0$, we obtain the eigenstates of the system under various disorder strengths by solving the Hamiltonian and then perform exponential fitting to extract $\xi$. The results are shown in \hyperref[Fig7]{Fig. 7(e)}. It is observed that $\xi$ decreases rapidly with increasing disorder strength. Interestingly, the energy distribution of $\xi$ is consistent with the unique density of states of the kagome lattice: the energies $E\approx-2$ and $E\approx0$ correspond to van Hove singularities, where $\xi$ is relatively large; whereas $E\approx2$ corresponds to the flat band, where the localization length is second only to that near the Fermi level. When $\Delta \ge 2$, $\xi < 3$, indicating that our calculations for the finite system with $L = 6$ are relatively reasonable.
Based on the combined behaviors of $N(0)$ and Anderson localization, we delineate a schematic shaded region in \hyperref[Fig1]{Fig.~1} to distinguish the correlated Anderson insulator from the disordered Mott insulator.

\begin{figure}[t]
	\centering
	\includegraphics[width=\linewidth]{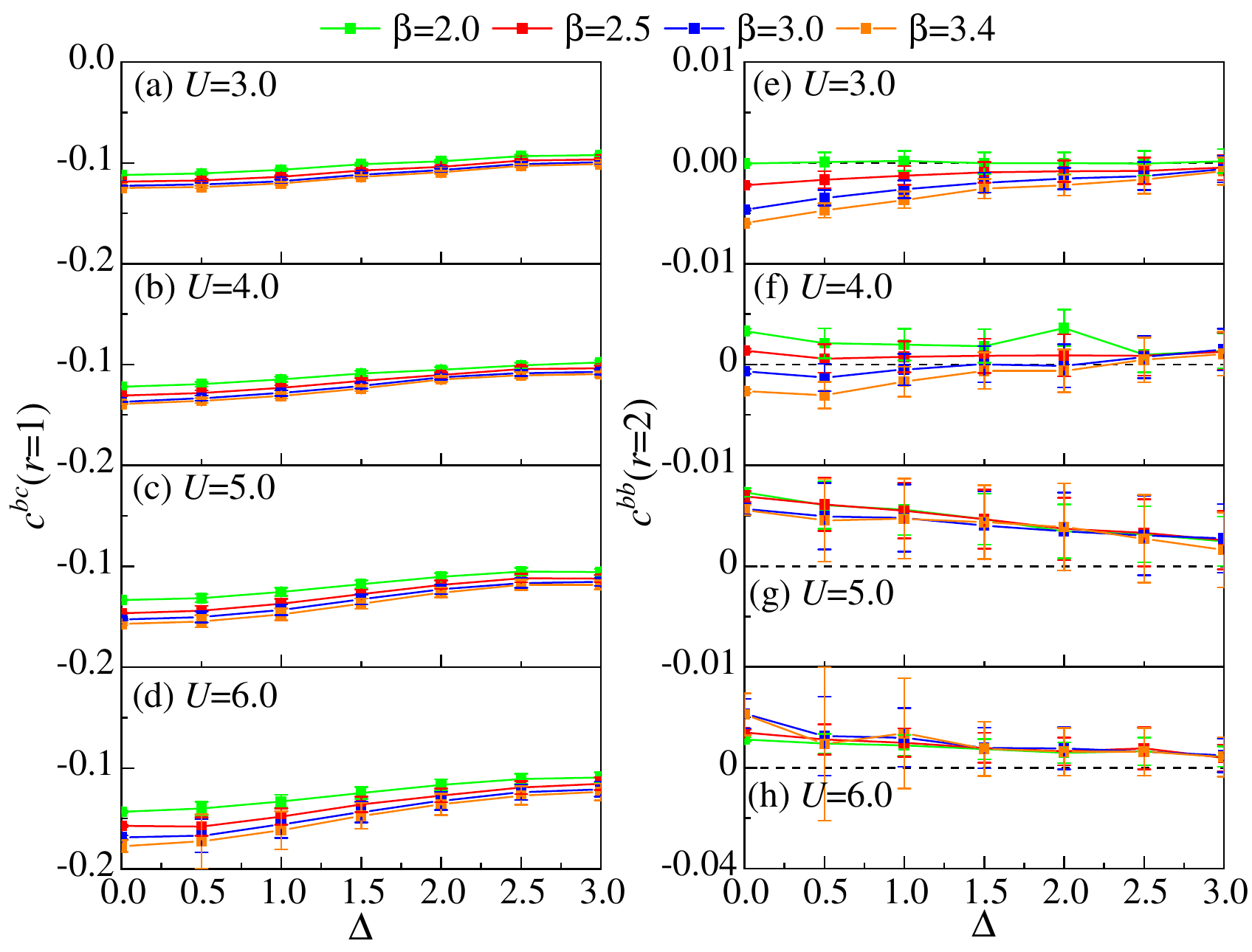}
	\caption{(a)-(d) Nearest-neighbor spin-spin correlation function between $b$ and $c$ sites $c^{bc}(r=1)$ and (e)-(h) next-nearest-neighbor correlation between $b$ sites $c^{bb}(r=2)$ as a function of disorder strength $\Delta$ for various correlation strength $U$ and temperature $T$.}
	\label{Fig8}
\end{figure}

To investigate the magnetic properties of the disordered kagome lattice, we examine $c^{bc}(r=1)$ as a function of $\Delta$ for various $U$ (See \hyperref[Fig8]{Fig. 8(a)-(d)}). For all values of $U$ and temperature $T$, $|c^{bc}(r=1)|$ exhibits a monotonic decrease from negative values with increasing $\Delta$. 
In Previous studies \cite{PhysRevB.76.144413}, the introduction of potential disorder in the kagome system relieves the frustration within the triangle due to the removal of the apical site, thereby enhancing local AFM correlations. Our study, however, introduces hopping disorder, which differs from the above understanding. 
Therefore, the effect of hopping disorder on the local AFM correlations in the kagome system is opposite to that of potential disorder.
At fixed $\Delta$ and $T$, $|c^{bc}(r=1)|$ grows with increasing $U$, indicating that stronger repulsive interactions promote electron localization and thereby enhance short-range correlations. 
$c^{bb}(r=2)$ is presented in \hyperref[Fig8]{Fig. 8(e)-(h)}. Notably, $c^{bb}(r=2)$ is suppressed by at least one order of magnitude relative to $c^{bc}(r=1)$. For $U=3.0$ and $U=4.0$, $c^{bb}(r=2)$ is nearly zero. For $U \geq 5.0$, $c^{bb}(r=2)$ gradually develops into weak ferromagnetic correlation. These findings demonstrate that the kagome lattice hosts strong local AFM correlation, which is partially suppressed by disorder. Based on the combined behavior of the conductivity and magnetic properties in the disordered correlated kagome system, we determined a schematic region separating the local AFM metal from the correlated Anderson insulator, as illustrated in \hyperref[Fig1]{Figure 1}.

\begin{figure}[t]
	\centering
	\includegraphics[width=\linewidth]{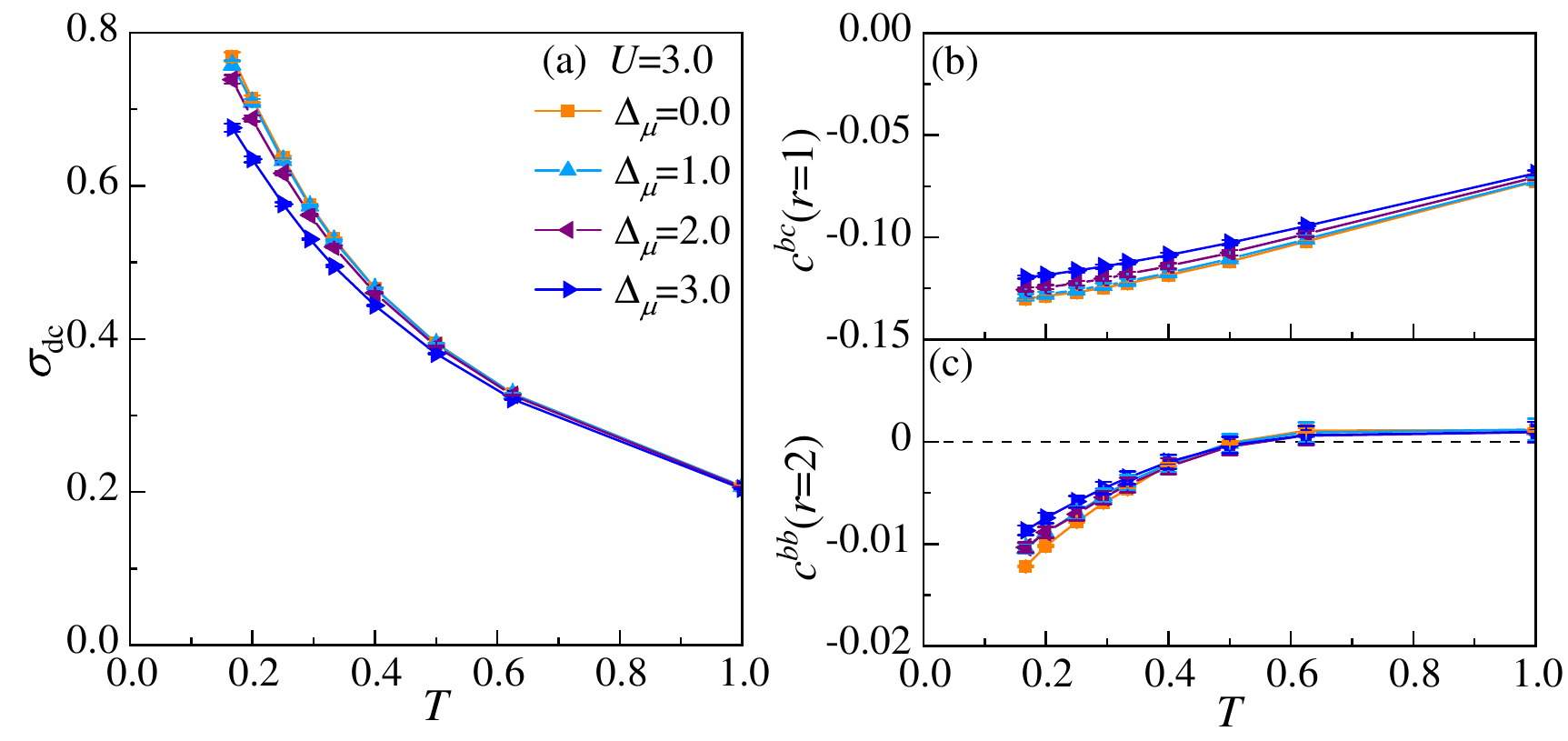}
	\caption{(a) dc conductivity $\sigma_{\mathrm{dc}}$ as a function of temperature $T$ for various disorder strength $\Delta_{\mu}$ at $U=3.0$. (b)-(c) Nearest-neighbor spin-spin correlation function between $b$ and $c$ sites $c^{bc}(r=1)$ and next-nearest-neighbor correlation between $b$ sites $c^{bb}(r=2)$ as a function of temperature $T$ for various disorder strength $\Delta_{\mu}$ at $U=3.0$. The legends in (b) and (c) are the same as that in (a).}
	\label{Fig9}
\end{figure}

Considering the introduction of both potential disorder and hopping disorder during substitutional doping, we investigate the impact of potential disorder on the MIT. The Hamiltonian is given as follows:
\begin{equation}
\hat{H} = -\sum_{\langle i,j \rangle,\sigma} t \left( \hat{c}_{i\sigma}^{\dagger}\hat{c}_{j\sigma}^{\phantom{\dagger}} + \hat{c}_{j\sigma}^{\dagger}\hat{c}_{i\sigma}^{\phantom{\dagger}} \right) - \sum_{i\sigma} \mu_{i} \hat{n}_{i\sigma} + U \sum_{i} \hat{n}_{i\uparrow}\hat{n}_{i\downarrow},
\label{Eq6}
\end{equation}
where $t=1$ and $\mu_{i}$ is randomly chosen from the interval $[\mu-\Delta_{\mu}/2,\, \mu+\Delta_{\mu}/2]$. \hyperref[Fig9]{Fig. 9(a)} presents $\sigma_{\mathrm{dc}}$ as a function of temperature for various $\Delta_{\mu}$ values at $U=3.0$. The results reveal that potential disorder suppresses $\sigma_{\mathrm{dc}}$ in a manner analogous to hopping disorder; however, no qualitative changes in transport properties are observed, even when $\Delta_{\mu}$ is increased to 3.0. We also examined the effect of potential disorder on magnetic correlations. \hyperref[Fig9]{Fig. 9(b)-(c)} display $c^{bc}(r=1)$ and $c^{bb}(r=2)$ as functions of temperature for different $\Delta_{\mu}$ values at $U=3.0$, respectively. It is evident that $\Delta_{\mu}$ weakly suppresses local magnetic correlations, a behavior consistent with that induced by hopping disorder.

\begin{figure}[t]
	\centering
	\includegraphics[width=\linewidth]{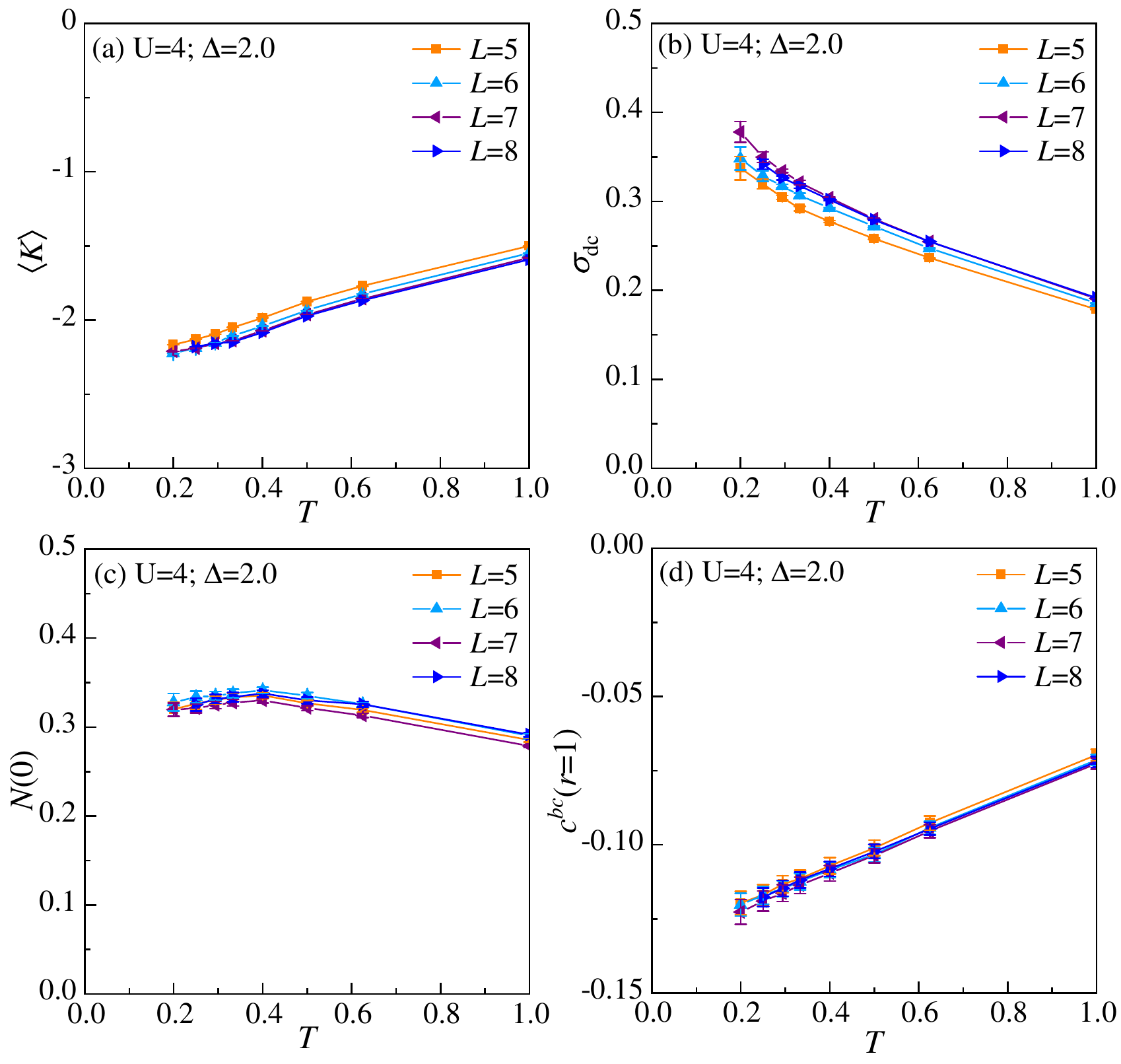}
	\caption{(a) Kinetic energy per site $\langle K \rangle$, (b) dc conductivity $\sigma_{\rm dc}$, (c) density of states at the Fermi level $N(0)$ and (d) spin-spin correlations on the nearest-neighbor $b$ and $c$ sites $c^{bc}(r = 1)$ as a function of temperature $T$ with the lattice size increasing from 5 to 8.}
	\label{Fig10}
\end{figure}

To ensure computational convergence, we examined the finite-size effects on $\langle K \rangle$, \(\sigma_{dc}\), $N(0)$ and \(c^{bc}(r = 1)\). \hyperref[Fig10]{Fig. 10} present the above quantities as a function of temperature \(T\) for different lattice sizes with $U=4.0$ and $\Delta=2.0$, respectively. The results indicate that all quantities exhibit negligible size dependence, with relative variations among different $L$ values remaining below $5\%$ across the entire ranges. The weak finite-size effect justifies the use of $L=6$ as a reliable representative of the thermodynamic limit in our main calculations \cite{PhysRevB.105.045132}.

\vspace{-0.2cm}
\section{IV. CONCLUSION}
\vspace{-0.3cm}

Using the DQMC method, we investigate the electronic transport and magnetic properties of the disordered Hubbard model on the kagome lattice. In the disorder-free honeycomb lattice, the results obtained by the DQMC method \cite{PhysRevLett.120.116601} are consistent with those from various previous methods (higher resolution) \cite{srep00992,PhysRevX.3.031010,PhysRevB.72.085123,PhysRevB.92.045111}. Moreover, it has been found that raising the temperature causes amorphous monolayer carbon to lose intermediate-range order and become an insulator \cite{s41586-022}, suggesting that generalized hopping disorder can induce a MIT in honeycomb-like structures, thereby indirectly validating the results of the DQMC method \cite{PhysRevLett.120.116601}. On the other hand, our results for the disorder-free kagome lattice are consistent with previous studies \cite{PhysRevB.104.L121118,PhysRevB.93.245123,PhysRevB.107.035134}.

The calculated $\langle \mathrm{sign} \rangle$ demonstrates that disorder can effectively alleviate the sign problem in systems with strong interaction strengths. At $U=3$, disorder drives the transition from an AFM metal to a gapless Anderson insulator. As the Coulomb repulsion increases, $\Delta_{\mathrm{c}}$ required for the MIT decreases, indicating that the interplay between disorder and interactions facilitates more effective electron localization. On the other hand, within the disordered insulator, there is a transition from an Anderson-like gapless insulator to a Mott-like gapped insulator. Furthermore, while disorder can partially suppress the AFM order, it remains insufficient to trigger a magnetic transition. Finally, we construct a phase diagram parameterized by disorder strength and correlation strength, which roughly distinguishes between AFM metal, correlated Anderson insulator, and disordered Mott insulator.

Correlating the degree of disorder (DOD) of amorphous solids with their physical properties has long been a formidable challenge in materials science and condensed matter physics. To address this, direct imaging of amorphous monolayer carbon grown by laser-assisted deposition has enabled the resolution of atomic configurations and the quantification of DOD \cite{sciadv.1601821,s41586-020}. Furthermore, varying the growth temperature allows for facile tuning of both DOD and electrical conductivity in amorphous monolayer carbon films---a modest increase of merely $25^{\circ}\mathrm{C}$ suffices to drive the system from a state with medium-range order to an electrically insulator \cite{s41586-022}. This strategy of controlling disorder via growth temperature provides a viable paradigm for inducing insulating behavior in kagome-based systems. Notably, a resurgence of interest in investigating disorder effects in correlated systems using optical lattice experiments. These ultra-cold atomic systems enable precise control over disorder and coupling parameters, thereby facilitating direct comparison between experimental data and theoretical predictions \cite{PhysRevLett.114.083002,PhysRevLett.116.140401}. The results presented herein may offer useful insights for future cold atom experiments. Finally, the reduction in the critical interaction strength required for the MIT in the presence of disorder may have significant implications for the application of kagome materials in logic and optoelectronic devices.

\vspace{-0.2cm}
\section{ACKNOWLEDGMENTS}
\vspace{-0.3cm}
This work was supported by NSFC (12474218) and Beijing Natural Science Foundation (No. 1242022 and 1252022). The numerical simulations in this work were performed at the HSCC of Beijing Normal University.

\vspace{-0.2cm}
\section{Data availability}
\vspace{-0.3cm}
The data that support the findings of this article are openly available \cite{D_2026_18809365}.

\bibliography{ref}
\bibliographystyle{style}

\end{document}